# Transition to a Bose-Einstein condensate of excitons at sub-Kelvin temperatures

Kosuke Yoshioka,[1,4] Eunmi Chae[2,4†] & Makoto Kuwata-Gonokami[1,3,4]

[1]*Department of Physics, Graduate School of Science, The University of Tokyo, 7-3-1 Hongo, Bunkyo-ku, Tokyo, 113-0033, Japan;*

[2]*Department of Applied Physics, Graduate School of Engineering, The University of Tokyo, 7-3-1 Hongo, Bunkyo-ku, Tokyo, 113-8656, Japan;*

[3]*Photon Science Centre, The University of Tokyo, 7-3-1 Hongo, Bunkyo-ku, Tokyo, Japan;*

[4]*Core Research for Evolutional Science and Technology, Japan Science and Technology Agency, 4-1-8 Honcho, Kawaguchi-shi, Saitama, 332-0012, Japan.*

[†]Present address: Department of Physics, Harvard University, Cambridge, Massachusetts 02138, USA.

**Bose-Einstein condensation (BEC) is a quantum mechanical phenomenon directly linked to the quantum statistics of bosons. While cold atomic gases provide a new arena for exploring the nature of BEC[1–4], a long-term quest to confirm BEC of excitons, quasi-Bose particles formed as a bound state of an electron-hole pair, has been underway since its theoretical prediction in the 1960s.[5,6] Ensembles of electrons and holes are complex quantum systems with strong Coulomb correlations; thus, it is non-trivial whether nature chooses a form of exciton BEC. Various systems have been examined in bulk[7–9] and two-dimensional semiconductors[10,11] and also in exciton-photon hybrid systems[12–14]. Among them, the 1$s$ paraexciton state in a single crystal of $Cu_2O$ has been a prime candidate for realizing spontaneous build up of a three-dimensional BEC that is decoupled from radiation field. The large binding energy and long lifetime enable preparation of**

**cold excitons in thermal equilibrium with the lattice. However, collisional loss[15,16] severely limits the conditions for reaching BEC. Such a system with a large inelastic cross section is excluded in atomic BEC experiments, where a small inelastic scattering rate and efficient elastic scattering are necessary for evaporative cooling. Here we demonstrate that it is nevertheless possible to achieve BEC by cooling paraexcitons to sub-Kelvin temperatures in a cold phonon bath. Emission spectra from paraexcitons in a three-dimensional trap show an anomalous distribution in a threshold-like manner at the critical number of BEC expected for ideal bosons. Bosonic stimulated scattering into the condensate and collisional loss compete and limit the condensate to a fraction of about 1%. This observation adds a new class of experimentally accessible BEC and provides us opportunities for exploring a rich variety of matter phases of electron-hole ensembles.**

The 1$s$ paraexciton is the pure spin triplet state, so it is decoupled from the radiation field. In high-quality single crystals, excitons show very long lifetimes of more than hundreds of nanoseconds,[16,17] reaching thermal equilibrium with the lattice. The corpuscular nature of excitons is well visualized in emission spectra reflecting their thermal distribution. Since paraexcitons are light quasi-particles (effective mass $m_{ex}$ = 2.6$m_0$ [$m_0$ is the electron mass at rest][18]), BEC has been believed to be attainable at moderate density ($n \sim 10^{17}$ cm$^{-3}$) at the temperature of superfluid helium (2 K)[7–9]. However, a two-body inelastic collision process that hampers efficient accumulation and cooling prevents the realisation of exciton BEC in Cu$_2$O[15]. A recent paraexciton density evaluation with a wide dynamic range[16] showed that rather dilute ($n \sim 10^{14}$ cm$^{-3}$) paraexcitons increase sub-linearly with the excitation rate, so that the crystal (and excitons) should suffer significant heating in achieving a density of $10^{17}$ cm$^{-3}$. This effect is well known as an efficient collision-induced loss channel for 1s orthoexcitons (the second-lowest exciton level, with lifetimes of a few ns) .[15]

In atomic BEC experiments[1–4,19], s-wave elastic scattering cross sections (or equivalently, scattering lengths) are crucial parameters. The elastic scattering cross section $\sigma_{el}$ should be significantly larger than the inelastic cross section $\sigma_{inel}$ to obtain a quantum degenerate gas via efficient evaporative cooling. Only the atomic species that fulfil this condition are pre-selected. Despite the importance of scattering properties, little information on the scattering cross section of paraexcitons has been obtained until recently. Here, we summarize the characteristics of the 1s exciton in $Cu_2O$, whose Bohr radius is about 0.7 nm. When the temperature is lowered, the inelastic cross section diverges (i.e. the two-body collision-induced loss coefficient does not depend on temperature). Such behaviour is typical of quantum-mechanical inelastic collisions and is known as the inverse velocity law of s-wave inelastic scattering.[20] The measured value for paraexcitons is $\sigma_{inel}$ = 80 nm$^2$ at 5 K[16]. From the inelastic cross section, we estimate an elastic cross section of $\sigma_{el} \approx 30$ nm$^2$ (see Supplementary Information); quantum Monte Carlo calculations[21] show $\sigma_{el}$ = 50 nm$^2$. Since the inelastic cross section is inversely proportional to the square root of the temperature, and the elastic cross section is temperature independent, the inelastic collisional loss rate is always higher than the elastic scattering rate and therefore governs the properties of the system at low temperature. To achieve BEC, we must set the critical density as low as possible by lowering the exciton temperature. Inefficient elastic collisions cannot redistribute the excess energy generated in inelastic collisions. A phonon bath ensures the thermal equilibrium of the excitons. In principle, BEC is possible even when inelastic collisions dominate. Such a situation is inaccessible in atomic experiments, and thus the present system can offer a unique opportunity to study a new type of BEC.

Here we use a $^3$He refrigerator to cool the crystal as low as 278 mK. To avoid ballistic diffusion of excitons[22] and to accumulate excitons under quasi-continuous-wave (cw) feeding, we prepare a three-dimensional harmonic potential trap with an inhomogeneous applied strain[23,24] (see Methods). We design a sample holder with a built-in spring that blocks any incoming heat to apply static pressure to the crystal. The trap frequencies in the *xy* and *z* directions (Fig. 1a) are $\omega_{x,y} = 2\pi \times 17$ MHz and $\omega_z = 2\pi \times 25$ MHz, respectively (see Supplementary Information), to be compared with the lifetime of 300 ns (see Supplementary Information). The excitation continuous wave (cw) laser incoherently generates 1*s* orthoexcitons (12 meV above the 1*s* paraexciton level) via a phonon-assisted absorption process in a cigar-shaped volume around the bottom of the trap (Figs. 1b, 1c). The orthoexcitons convert into 1*s* paraexcitons in nanoseconds and subsequently flow towards the bottom of the trap potential. Paraexcitons release their excess kinetic energy to the phonon bath via scattering with acoustic phonons on their way to the bottom; thus, sufficiently cold paraexcitons accumulate near the potential minima. We detected spatially resolved, time-integrated luminescence spectra of direct radiative recombination of paraexcitons, which is slightly allowed by the strain field.[25]

First, we examined the temperature dependence of the signal (Figs. 2a–2c) under low-density excitation ($3 \times 10^6$ paraexcitons in the trap at the lowest temperature), where density-dependent processes are negligible. Above 1 K, the size of the cloud (Fig. 2d) and the spectral width (Fig. 2e) coincide perfectly with a theoretical prediction assuming the classical thermal distribution (modified for direct luminescence[23]). This ensures that the exciton gas is in thermal equilibrium with the lattice. However, at $T_L \leq$ 1 K, the exciton temperature remains higher than that of the lattice, indicating that exciton cooling by phonon emission becomes inefficient; namely, the exciton-photon scattering time becomes comparable to or longer than the exciton lifetime. From the spatial and spectral widths, the temperature of the exciton gas stays at $T = 0.8$ K when

$T_L \leq 0.6$ K. The critical number for BEC of non-interacting Bose gas at $T = 0.8$ K is $N_c = 7.6 \times 10^8$ from the equation

$$N_c = \varsigma(3)\left(\frac{kT}{\hbar\overline{\omega}}\right)^3, \tag{1}$$

where $\varsigma(x)$ and $\overline{\omega}$ denote the Riemann zeta function and geometric angular frequency ($\overline{\omega} = \sqrt[3]{\omega_x \omega_y \omega_z} = 2\pi \times 19$ MHz), respectively. The increase in integrated intensity (Fig. 2f) at lower $T_L$ reflects the increased capture efficiency of the trap and increased emission collection of the shrunken cloud image on the entrance slit of the spectrometer (7.5 μm wide on the cloud).

Note that two distinct features are needed to pursue paraexciton BEC at sub-Kelvin temperatures. The first is coherent harmonic motion of paraexcitons in a trap. To establish the energy structure of excitons in a three-dimensional harmonic potential trap, their mean scattering time should be longer than the period of harmonic oscillation. Therefore, the exciton-phonon scattering rate ($\Gamma_{phonon}$) must be less than the trap frequency ($f_{trap}$): $\Gamma_{phonon} < f_{trap}$. This has not been realized in previous trap experiments at liquid $^4$He temperatures, where $\Gamma_{phonon}$ is a few hundred MHz and $f_{trap}$ is 13 MHz.[23] At sub-Kelvin temperatures, the phonon scattering time is reduced and is comparable to the exciton lifetime of 300 ns (see Supplementary Information). This implies that $\Gamma_{phonon}$ is a few MHz, which satisfies the condition described above. The second is smaller heating resulting from weaker excitation power. Owing to the large $\sigma_{inel}$, the collision-limited paraexciton number $N$ effectively scales as $N \propto G^{0.5}$, where $G$ is the generation rate of excitons or the optical excitation power. Because the exciton gas is captured in a three-dimensional harmonic potential, the critical exciton number $N_c$ scales as $N_c \propto T^3$ ($N_c \propto T^{3/2}$ in free space). Therefore, the optical power required to prepare the critical number in a harmonic trap is proportional to $T^6$, whereas the specific heat of semiconductors scales as $T^3$. This naive consideration indicates that the lower the exciton temperature is, the more favourable the conditions are for achieving BEC with the least lattice temperature increase.

To increase the number of trapped paraexcitons while keeping the lattice temperature unchanged ($T_L$ = 354 mK), we chop the pump beam with appropriate duty cycles to make the average power constant (see Methods). For ideal Bose particles well above the critical number, a large occupation number of the ground state is expected. Instead, as shown in Fig. 3, we have observed anomalous spectra when the paraexciton number crosses the critical number. With increasing estimated paraexciton number, $N = 2 \times 10^7$, $5 \times 10^8$ and $2 \times 10^9$ (Figs. 3a–3c, respectively), a high-energy signal appears suddenly up to around 400 μeV above the potential minima. In Fig. 3d, we plot the integrated emission intensity ratio between the higher-energy part (2.01924–2.02004 eV) and the bottom part (2.01902–2.01921 eV) versus the paraexciton number. A threshold-like rise of the ratio near $N_c$ at an exciton temperature of 0.8 K is apparent, indicating that the BEC transition occurs at the condition expected for ideal bosons. We also examine the temperature dependence measurements under a constant excitation density. We set the excitation level at the maximum density of Fig. 3d, $N = 2 \times 10^9$ at $T_L$ = 354 mK. The lattice temperature dependence of the ratio also shows an abrupt increase at $T_L \sim 400$ mK (Fig. 3e), demonstrating the BEC transition of excitons (lowering $T_L$ results in a slight exciton temperature drop and an increased exciton number). In estimating $N$, we first calculated $N$ in the weakest excitation linear regime from the amount of optical absorption (36% of incident power on the refrigerator) and the lifetime of paraexcitons, assuming an overall trap collection efficiency of 30%. The validity of this method of exciton counting has been confirmed by exciton Lyman spectroscopy.[16] Using data from this low-density regime, we determined $N$ for higher-density excitations by comparing the emission signal intensity of the bottom part.

As discussed above, in this case inelastic collision is essential for inducing the anomalous spatial spectral profiles of the signal. Therefore, we exclude the effect of the renormalized energy due to elastic scattering of the condensate.[26] We propose a feasible scenario: competition between bosonic stimulated scattering and inelastic scattering-induced loss in the condensate. Considering the inelastic scattering cross section (or collisional loss coefficient on the order of $10^{-16}$ cm$^3$/ns) and the spatial spread of the ground state wavefunction of 1.4 µm at full width at half maximum (FWHM), the ground state is apparently unstable against ~$10^9$ condensed paraexcitons. This yields a collision-induced lifetime of less than 100 fs. By analysing rate equations considering collisional loss in the condensate and its redistribution to the thermal part, we estimate that the maximum condensate fraction is around 1% (80% in the case of ideal bosons) with a bosonic stimulated scattering rate of 7 ns$^{-1}$. Paraexcitons expelled from the condensate explains the enhancement of the observed signal in the high energy part (see Supplemenary Information). This situation is analogous to several examples of atomic condensates. The condensate fraction for atomic hydrogen is several percent due to the small $s$-wave elastic scattering length and three-body collisional loss.[4] A 'relaxation explosion' is even discussed.[28] For $^7$Li atoms with attractive interactions, the observed condensate fraction is a few percent at most, as determined by the attractive forces and the repulsion due to position-momentum uncertainty.[3,27]

To prepare a large condensate fraction, we must reduce the local density below $10^{15}$ cm$^{-3}$. Therefore, making colder exciton gas is essential. Active methods to enhance exciton-phonon coupling, together with better sample quality, are desired. A dynamic trap can be helpful for making a shallower trap potential upon growth of the condensate. Because inelastic scattering dominates elastic scattering, evaporative cooling does not work in this system. Elucidating the mechanism for the large inelastic scattering cross section is apparently crucial. It is interesting to compare our results with recently observed ultracold chemical reactions of molecules[29] and rapid inelastic collision of quantum degenerate *p*-wave molecules.[30]

**Supplementary Information** accompanies the paper on **www.nature.com/naturephysics**.

**Acknowledgements** We thank A. Mysyrowicz for discussions and long-term collaboration and N. Naka for discussions on trapping excitons. This work was supported by Grant-in-Aid for Scientific Research on Innovative Area 'Optical science of dynamically correlated electrons (DYCE)' 20104002.

**Author Contributions** K.Y. conducted the experiment and analysed the data. E.C. helped make the strain-induced trap for paraexcitons. K.Y. and M.K.-G. wrote the paper. M.K.-G. planned the project.

**Author Information** The authors declare no competing financial interests. Correspondence and requests for materials should be addressed to M.K-G. (gonokami@phys.s.u-tokyo.ac.jp).

**Figure 1 Capturing cold 1*s* paraexcitons in $Cu_2O$ single crystal. a,** Inhomogeneous strain field is formed by pressing the single crystal with a glass lens. The field, located 133 $\mu$m below the surface of the crystal, acts as a three-dimensional harmonic trap for paraexcitons. A focused excitation laser beam passes through the bottom of the trap. **b, c,** The strain field causes a position-dependent energy shift for both ortho- and paraexcitons. The excitation beam creates orthoexcitons only around the trap via a phonon-assisted absorption process; they rapidly convert to paraexcitons with their position almost unchanged. Paraexcitons flow into the bottom of the trap, dissipating their excess energy to the lattice. Accumulated paraexcitons move along the trap potential coherently owing to the very small interaction rate of cold paraexcitons with phonons.

**Figure 2 Coherent trapping of 1*s* paraexcitons at sub-Kelvin lattice temperatures.** Trapped paraexciton gas is observed via direct luminescence, which is only weakly allowed under the applied strain field. The spatially resolved luminescence is recorded as a function of $T_L$ under the weakest excitation condition. **a,** $T_L$ = 1.5 K. **b,** $T_L$ = 800 mK. **c,** $T_L$ = 287 mK. Dashed curves are calculated potential profiles of the harmonic trap. **d–f,** Spatial FWHM, spectral FWHM and total luminescence intensity of the paraexciton gas along the *z*-axis, respectively, versus the lattice temperature. Dashed curves are theoretical predictions assuming the classical thermal distribution. Green lines show experimental resolution of our setup.

**Figure 3 Anomalous luminescence of 1*s* paraexcitons when crossing the phase boundary for Bose-Einstein condensation.** Spatially resolved luminescence signals taken at $T_L$ = 354 mK, where the estimated paraexciton number *N* in the trap is **a,** $2 \times 10^7$ **b,** $5 \times 10^8$ and **c.** $2 \times 10^9$. Signal intensities are normalized. Arrows show the boundary of the higher-energy (2.01924–2.02004 eV) and bottom (2.01902–2.01921 eV) parts of the signal. **d,** Ratio between the higher-energy and bottom parts of the signal versus the estimated paraexciton number. Phase space density is calculated from the estimated number and the volume of the crowd at exciton temperature $T$ = 0.8 K, assuming ideal, non-interacting bosons. **e,** The ratio versus the lattice temperature under a constant excitation density ($N$ = $2 \times 10^9$ at 354 mK).

**Methods**

A cryogen-free $^3$He refrigerator (Oxford Instruments, Heliox-ACV) was used to cool the high-purity $Cu_2O$ single crystal conductively. The size of the optical windows and the numerical aperture were carefully chosen to prevent any heat load into the coldest plate of the refrigerator. The lowest crystal temperature without irradiation of the excitation beam was 278 mK. The sample was $5.3 \times 5.3 \times 9.1$ mm in size. Stress was applied along the [100] axis of the crystal by pressing the crystal with a lens and a built-in spring. Weak luminescence light from the trapped paraexciton gas was collected from the (110) surface. The excitation laser was a continuous-wave (cw) ring dye laser at 606 nm (Coherent Inc., 899-21). For an incident optical power of more than 250 µW, the excitation beam was chopped using an acousto-optic modulator and a high-isolation switch and was coupled into a single-mode optical fibre to reduce scattered light. We thereby obtained a high on-off extinction ratio of $2 \times 10^7$. The repetition rate was 500 Hz. We set the duty cycle $D_{cycle}$ when increasing the 'on' power $P_{on}$ of the quasi-cw excitation light so that the average power is constant. However, at high excitation intensity, the paraexciton number increases only sub-linearly. This results in a small integrated luminescence intensity because the duty cycle is reduced. In such cases, a high extinction ratio is essential because weak residual light in the 'off' time results in a considerable amount of signal under weak excitation. The magnified ($\times 4$) luminescence image of the trapped exciton gas was collected (numerical aperture 0.07) and dispersed by a 50-cm spectrometer. An EMCCD camera (Andor, DU970N-BV) detected the time-integrated signal. The integration time ranged from 100 s to 1500 s. The entrance slit of the spectrometer, which was parallel to the *z*-axis, captured only the central 7.5 µm (30 µm in the magnified image) of the gas. The vibration (amplitude ~10 µm) of the sample was actively cancelled using a pair of PZTs in the imaging section of the setup by monitoring the displacement using a laser diode and a position-sensitive photodetector.

The signal intensity of the $\Gamma_5^-$-phonon-assisted luminescence of paraexcitons was comparable to the detection noise level in our allowable integration time.

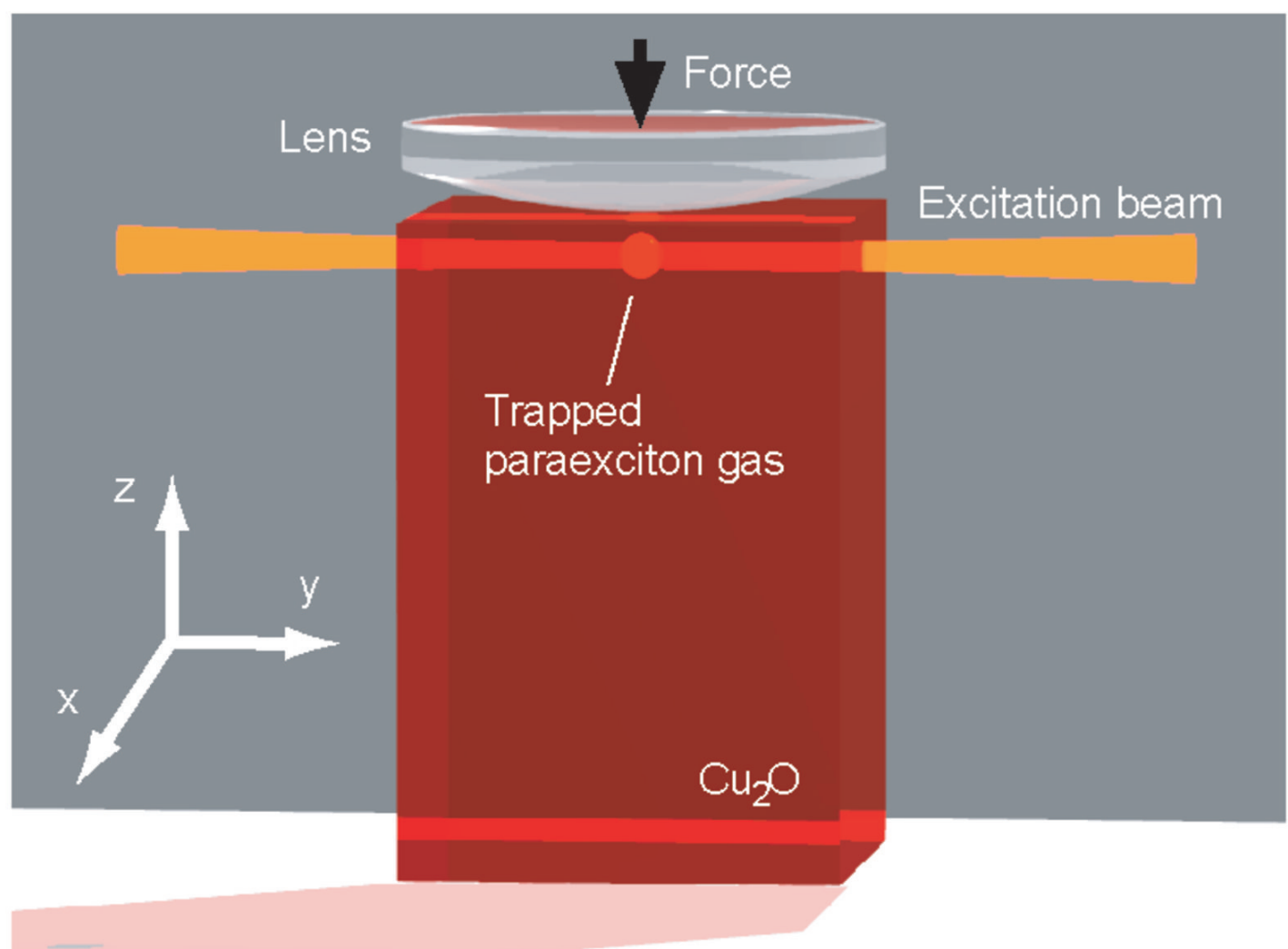

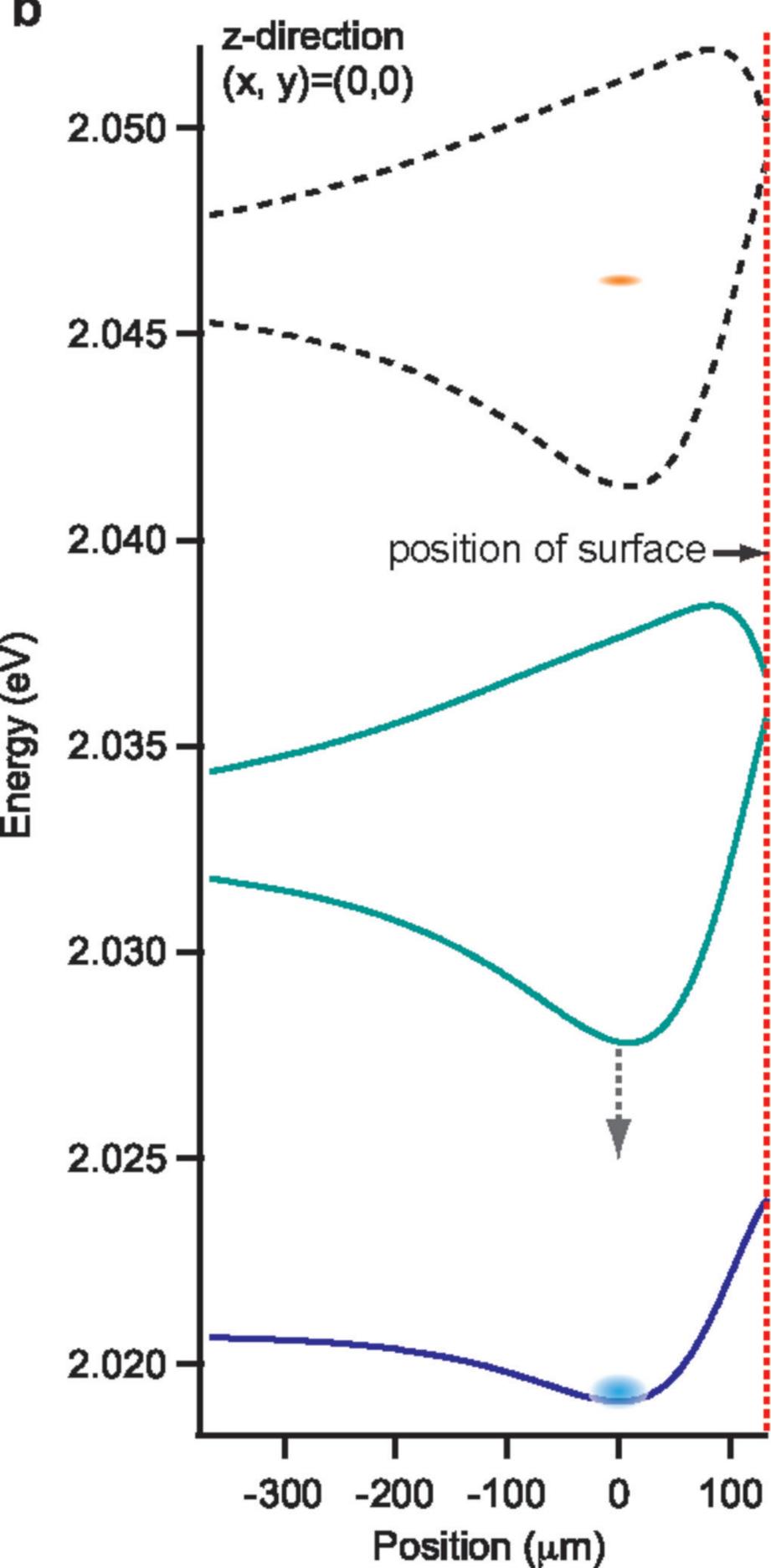

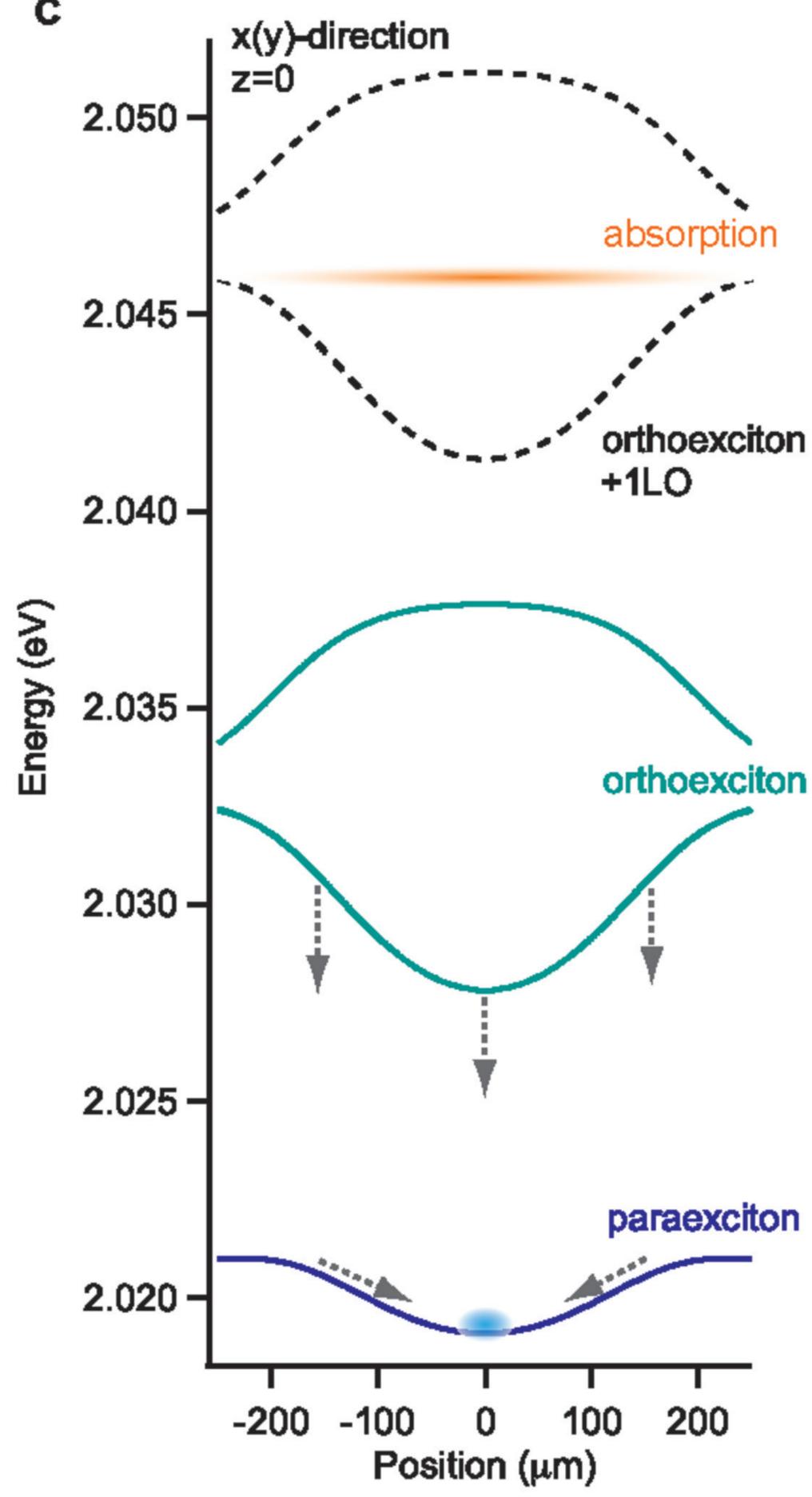

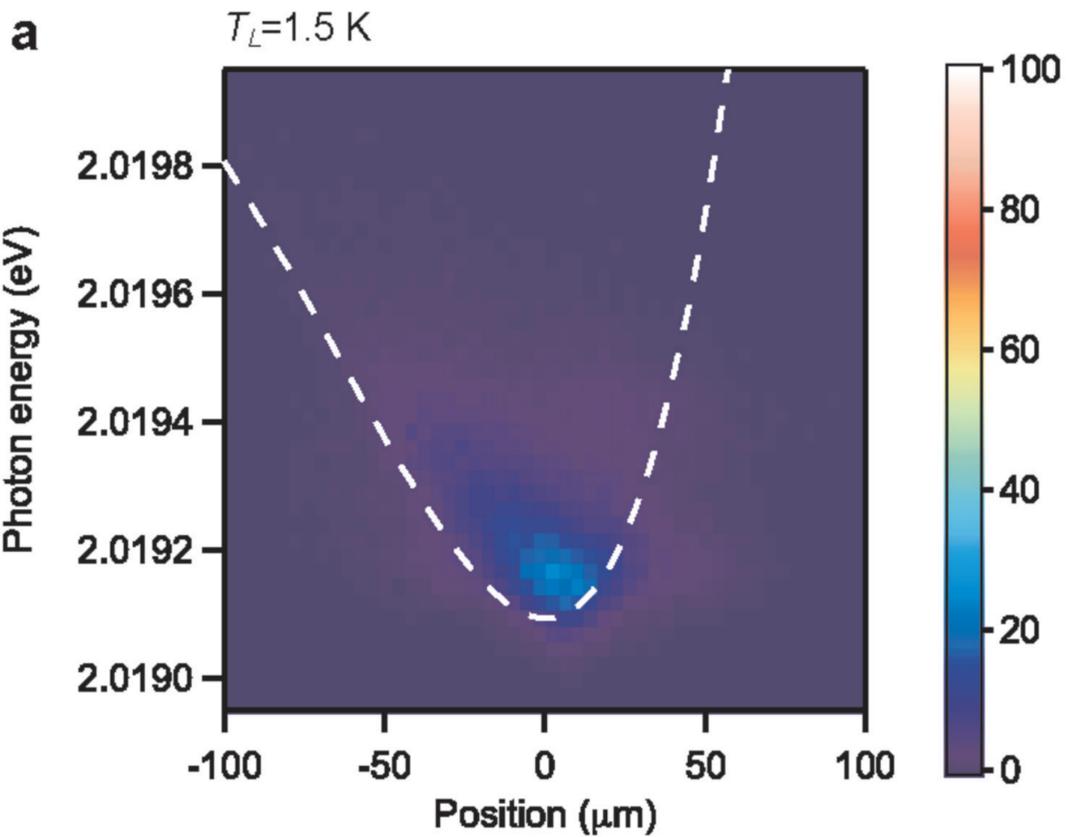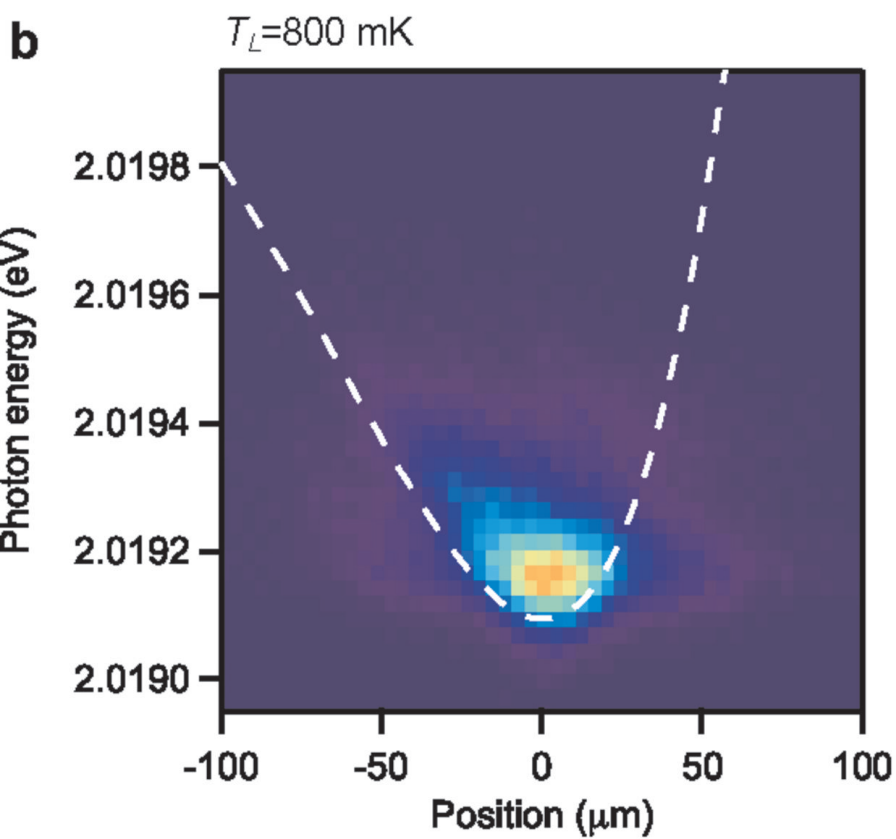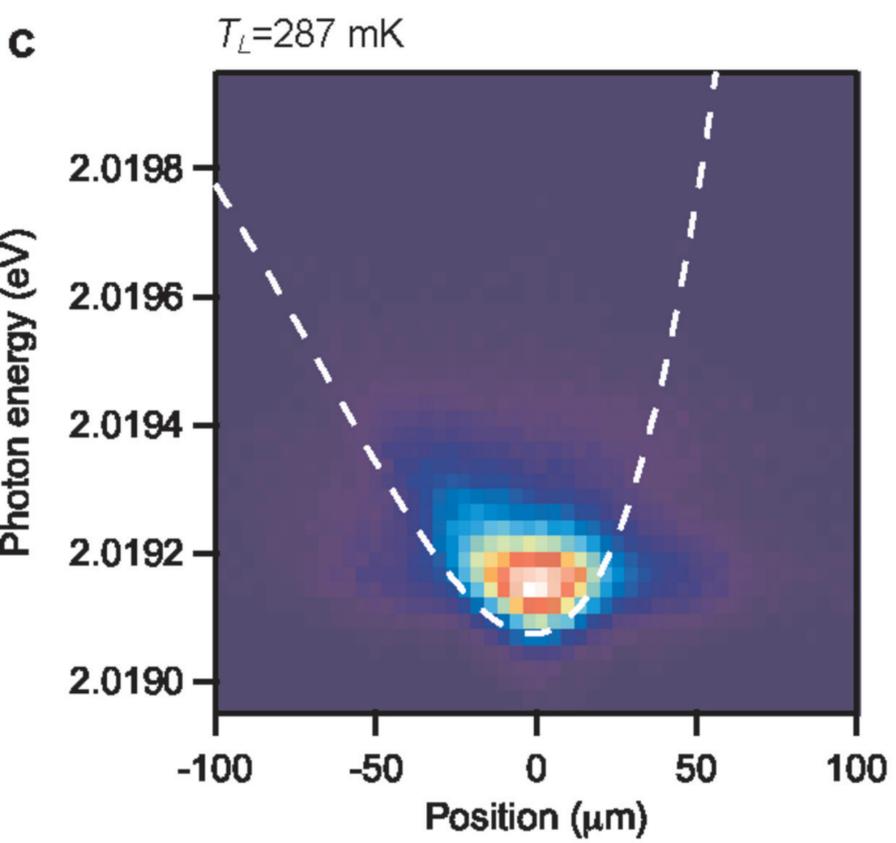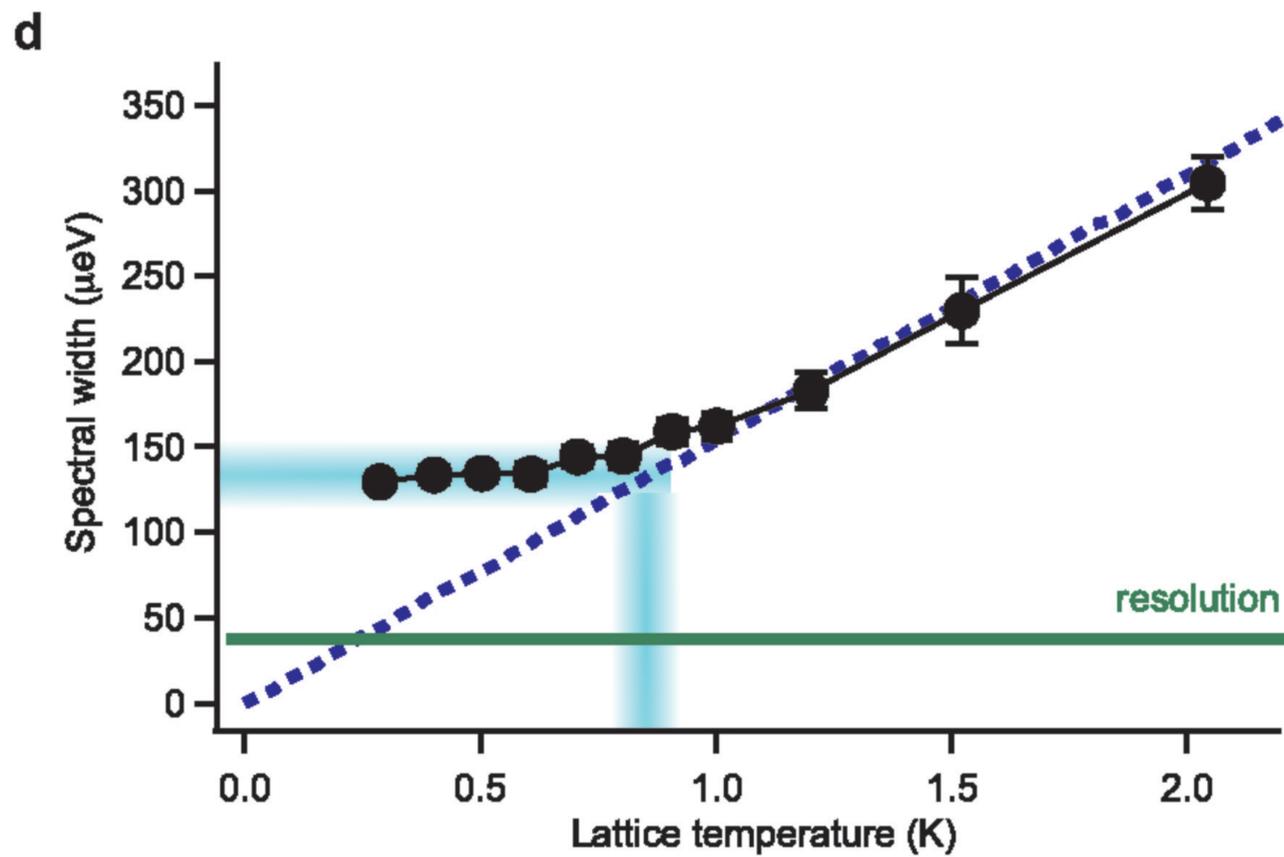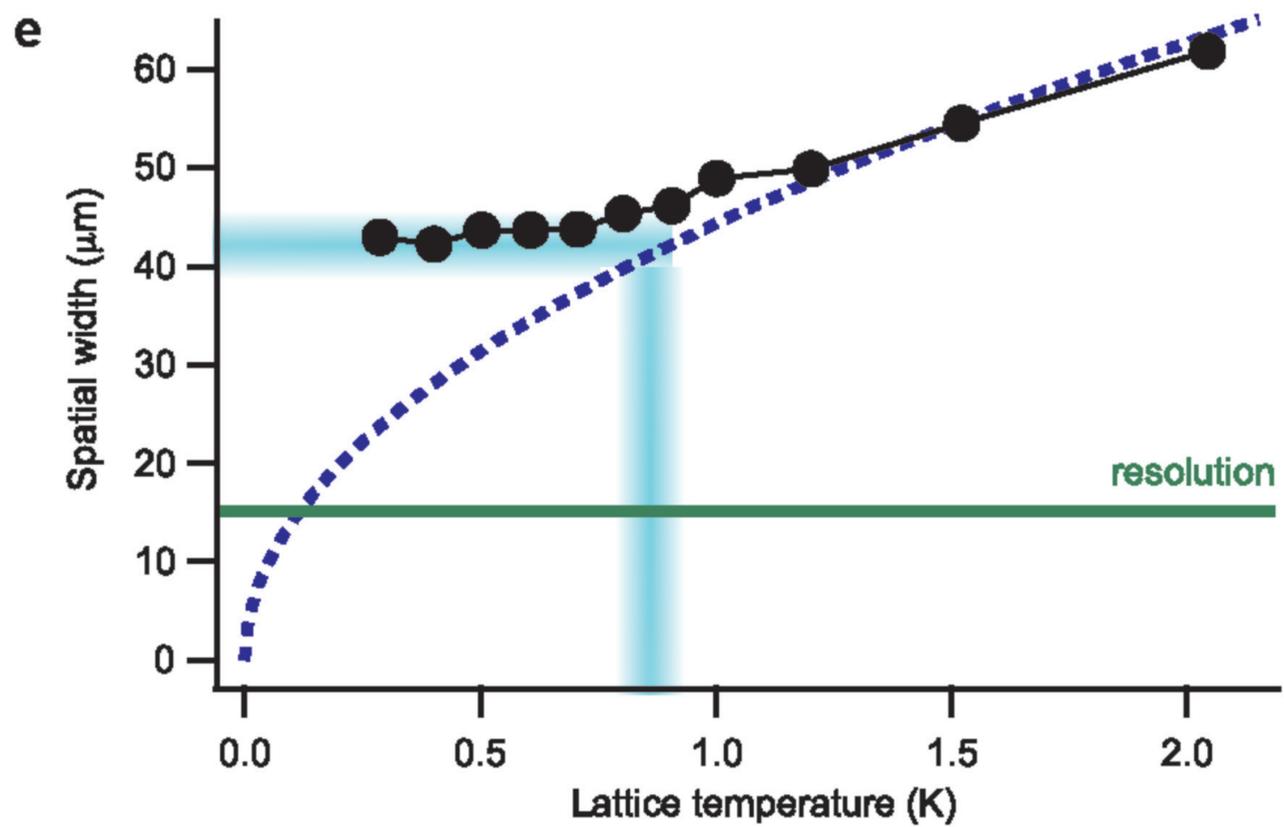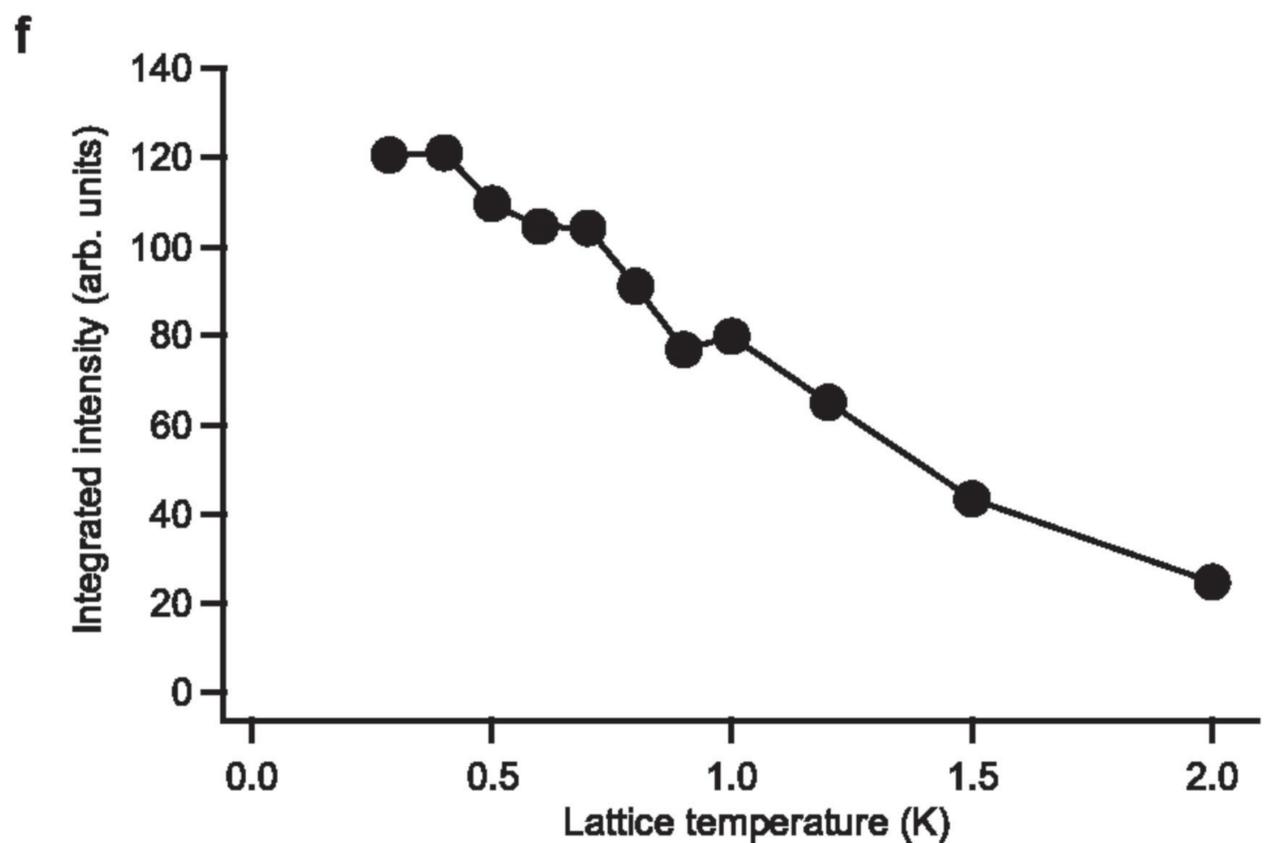

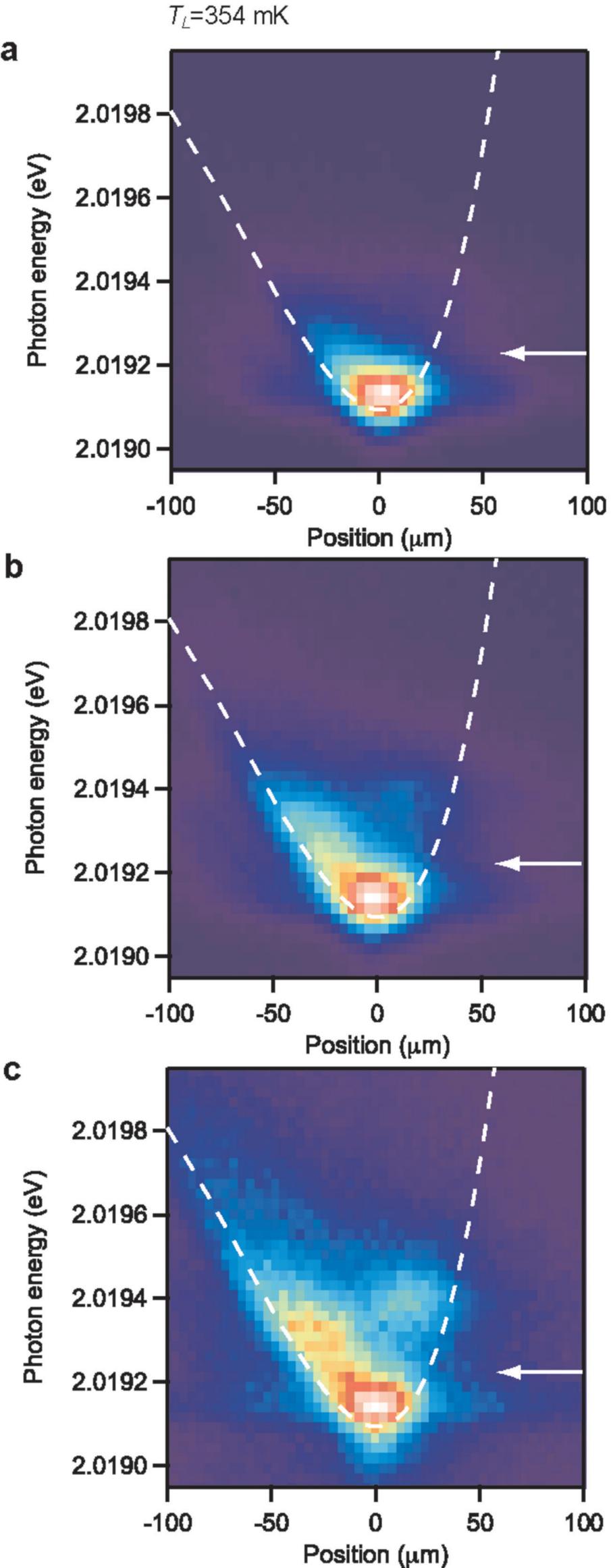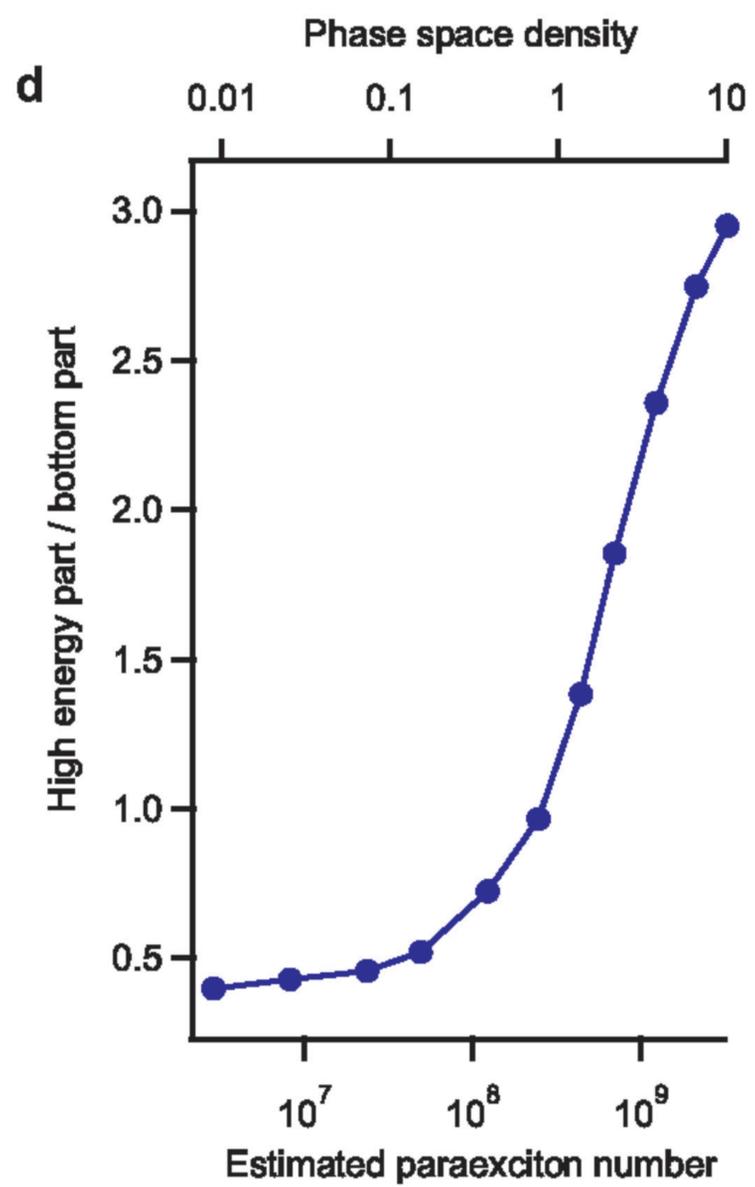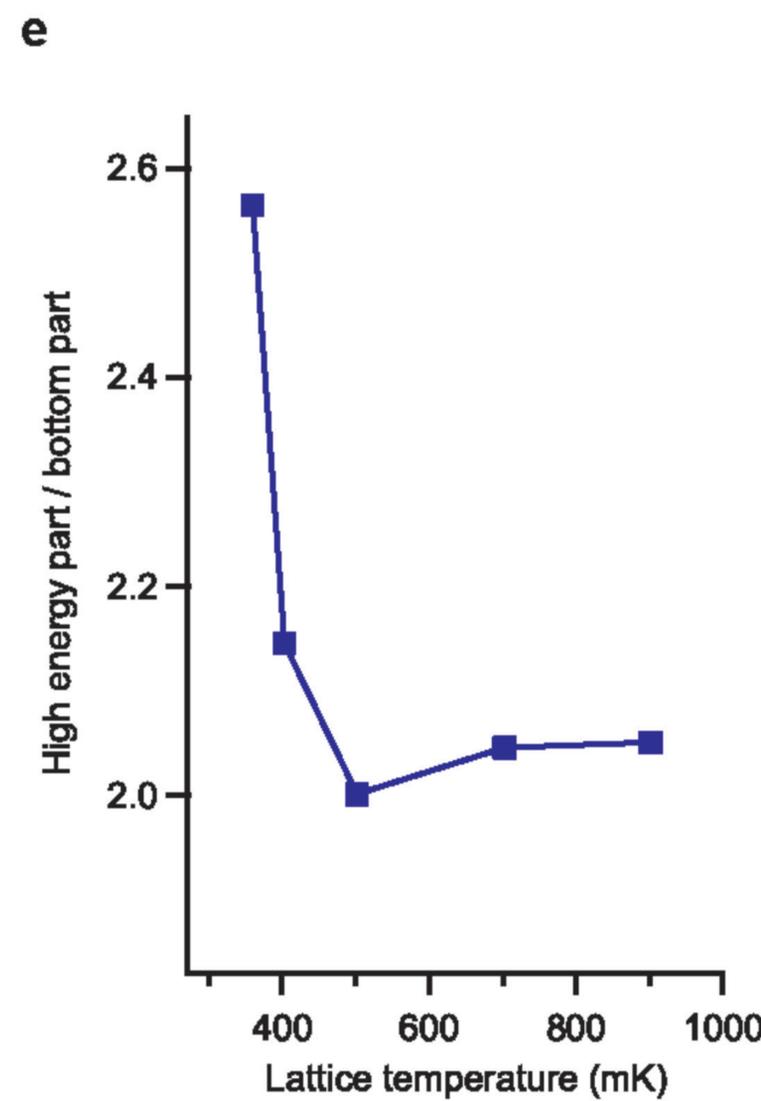

**Supplementary Information**

**Kosuke Yoshioka *et al.*.**

**Observation of the trap potential for 1*s* paraexcitons**

In addition to a theoretical analysis based on the well-established Hertzian contact problem,[1] we have also experimentally determined the trap potential. The procedure is as follows: We positioned the beam 220 μm away from the bottom of the trap potential. A small portion of the 1*s* paraexciton gas created at the excitation position (via a down-conversion process from 1*s* orthoexcitons) emits photons with the corresponding wavelength while flowing into the bottom of the trap. Supplementary Fig. 1a, taken at $T_L = 293$ mK, is a typical result observed along the *z*-axis under weak excitation with no density-dependent loss of the paraexciton gas. By making this comparison, we can also determine the potential curves along the *x*- and *y*-axes. The potential is harmonic for exciton gases in the temperature range considered in our experiment, except for an anomalous distribution at strong excitation where the thermal component of the gas experiences an asymmetric potential in the *z*-direction.

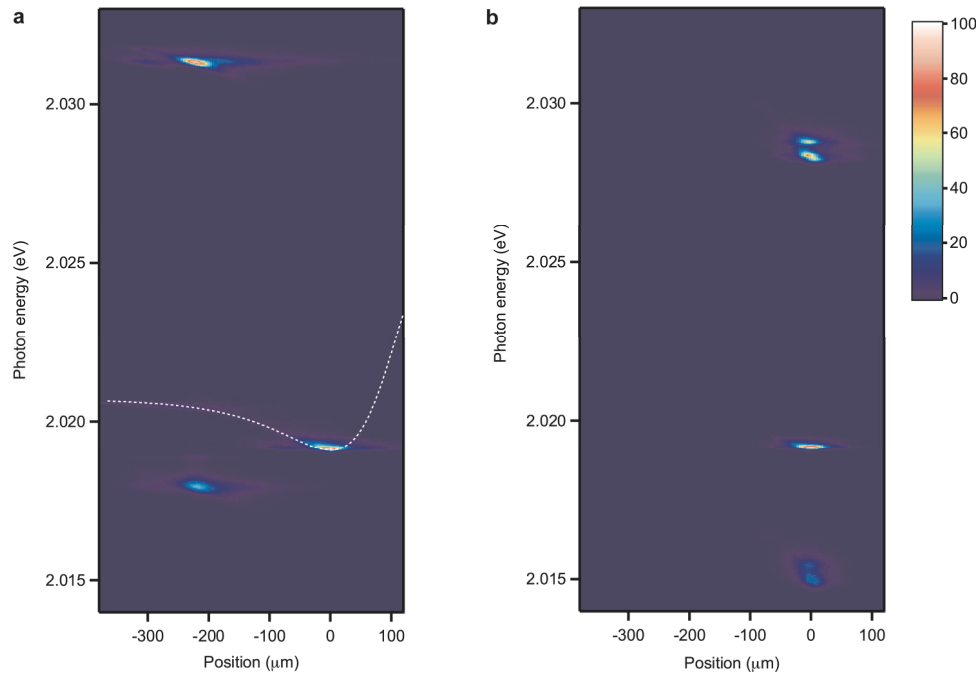

Supplementary Figure 1. Spatially resolved luminescence spectra at $T_L$ = 293 mK. **a.** Excitation beam was positioned 220 μm away (along the *z*-axis) from the potential minima for paraexcitons. While flowing into the minima, a tiny portion of the paraexciton gas recombined and emitted photons. We can thus trace the potential curve for paraexcitons. Dotted curve shows the calculated potential, which agrees well with the experimental curve. Signals around 2.032 eV and 2.018 eV correspond to the direct and LO-phonon assisted luminescence of 1*s* orthoexcitons. **b.** Beam position is set at the potential minima. All data shown in the main text were taken in this configuration.

**Lifetime of 1*s* paraexcitons in our three-dimensional trap**

We observed time-resolved, spatially resolved luminescence spectra of paraexcitons using a CCD camera with a gated image intensifier after passing the signal through a 50-cm spectrometer. The time resolution was 100 ns. Supplementary Fig. 2 shows the time-resolved intensity at $N_{max} = 3 \times 10^9$ and $T_L$ = 350 mK. After a rapid initial decay due to collision-induced loss, we observed a slow decay with a single exponential decay constant, $\tau$ = 300 ns, which we take as the lifetime of paraexcitons.

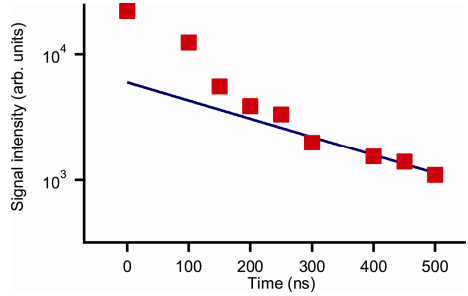

Supplementary Figure 2. Time-resolved, spatially integrated intensity of the paraexciton luminescence at 350 mK (red squares). Maximum estimated number is $3 \times 10^9$. Time resolution is 100 ns. Solid line is the theoretical curve with a lifetime of 300 ns.

**Signal intensities versus incident power**

In Supplementary Fig. 3, we show the signal intensity of the bottom part of the trapped paraexcitons (left-hand side of the vertical axis) and the ratio (right-hand side of the vertical axis) as a function of the optical power on the incident window of the $^3$He refrigerator. Measurements showed that 36% of the incident power was absorbed by phonon-assisted absorption of 1$s$ orthoexcitons. Here the Fresnel reflection losses at the interfaces of the windows and the crystal, as well as the scattering loss due to crystal imperfections, were taken into account by measuring the transmission under the far-off resonance condition. The signal intensity $I_{sig}$ was measured from the obtained signal intensity $I_{obt}$ using the expression

$$I_{sig} = \frac{I_{obt}}{RT_{integ}},$$

where $R$ is the duty cycle of the excitation light and $T_{integ}$ is the integration time. By doing this, we normalized the signal intensity for various excitation conditions and integration times. At the lowest-intensity condition in our experiment (10.5 μW), we could safely neglect the collision-induced loss of paraexcitons. In this condition, we estimated the number of trapped paraexcitons to be $3\times10^6$ from the measured lifetime of 300 ns, assuming a 30% collection efficiency to the trap. We estimated the paraexciton number under more intense excitation by comparing the signal intensities of the bottom part with the corresponding intensity under the lowest-intensity condition discussed above.

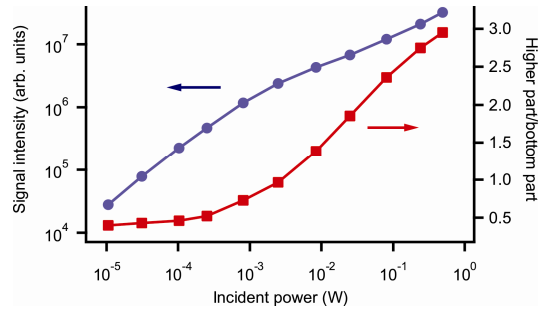

Supplementary Figure 3. Signal intensity of the bottom part (blue circles, left) and the ratio between the higher and bottom part (red squares, right) versus the incident power of the $^3$He refrigerator.

**Estimation of the condensate fraction and the bosonic stimulated scattering rate**

The anomalous energy distribution is due to inelastic collision-induced loss in the condensate, although collision loss is also present for the thermal part. To explain the threshold-like behaviour of the anomalous distribution, the number lost and redistributed per second should be much larger for the condensate than for the thermal part. Under this condition, we can estimate the number of thermal and condensate excitons based on the following rate equations. They model collision-induced loss and spontaneous bosonic-stimulated scattering into the condensate, as well as the lifetime of paraexcitons in the dilute limit.

$$\dot{N}_{th} = -\frac{1}{\tau} N_{th} - A \frac{N_{th}}{V} N_{th} - \gamma N_c N_{th} + \frac{1}{2} A \frac{N_{th}}{V} N_{th} + \frac{1}{2} A \frac{N_c}{\alpha V} N_c, \qquad (1)$$

$$\dot{N}_c = -\frac{1}{\tau} N_c + \gamma N_c N_{th} - A \frac{N_c}{\alpha V} N_c, \qquad (2)$$

where $\tau$ is the lifetime in the dilute limit, $A$ is the collision-induced loss coefficient ($A \sim 10^{-16}$ cm$^3$/ns), and $N_{th}$ and $N_c$ are the number of thermal and condensate paraexcitons, respectively. The rate $\gamma N_c$ is due to bosonic-stimulated scattering into the condensate. $V$ denotes the exciton volume of the thermal gas and $\alpha V$ is the condensate volume ($\alpha$ is the ratio). In two-body collisional loss for both the thermal and condensate parts, the remnant exciton is assumed to return to the thermal part. The last term in Eq. (1) yields the anomalous high-energy component. The condition for obtaining the anomalous distribution, as we observed in the steady state, is

$$\frac{1}{2} A \frac{N_{th}}{V} N_{th} \ll \frac{1}{2} A \frac{N_c}{\alpha V} N_c. \qquad (3)$$

Thus, the condition for the condensate fraction can be written as

$$\frac{N_c}{N_{th}} \gg \sqrt{\alpha}. \qquad (4)$$

The term $\sqrt{\alpha}$ ranges from $5.7 \times 10^{-3}$ (in the weakest-excitation case) to $1.4 \times 10^{-3}$ (in the strongest-excitation case), providing the lower limit of the condensate fraction. If the condensate fraction is close to unity, the luminescence intensity should decrease remarkably because no luminescence can be expected from the condensate owing to the momentum selection rule in the optical emission process. However, as shown in Supplementary Fig. 3, no significant decrease appears in the total signal intensity, so that it yields the upper limit of the condensate fraction. As a result, we estimate the condensate fraction to be around 1%. We can also roughly estimate the bosonic-stimulated scattering rate to the condensate from Eq. (2). At high density, the first term on the right-hand side can be omitted. Therefore, the steady state condition results in the relation

$$\gamma = \frac{A}{\alpha V} \frac{N_c}{N_{th}}. \tag{5}$$

For a typical set of parameters above the critical number at 0.8 K, we obtain $\gamma = 700$ s$^{-1}$ (the corresponding scattering rate is $\gamma N_c = 7$ ns$^{-1}$).

**Estimation of the elastic scattering cross section**

On the basis of a simple discussion of quantum-mechanical inelastic scattering of slow particles,[2] we can estimate the magnitude of the elastic scattering coefficient of paraexcitons from the inelastic one. In the case of $s$-wave scattering, the scattering matrix $S_0$ can be written simply as

$$S_0 \approx 1 - 2ik\alpha, \tag{6}$$

where $\alpha = \alpha' + i\alpha''$ is the complex scattering length, and $k$ is the wavenumber. The elastic ($\sigma_{el}$) and inelastic ($\sigma_{inel}$) cross sections are expressed as

$$\sigma_{el} = 4\pi|\alpha|^2, \tag{7}$$

$$\sigma_{inel} = \frac{4\pi|\alpha''|}{K_{ex}}, \tag{8}$$

where $K_{ex}$ is the thermal wavenumber of the paraexcitons. According to a measurement of the inelastic scattering cross section using the excitonic Lyman-$\alpha$ transition[3], $\sigma_{inel}$ = 85 nm$^2$ at 5 K ($K_{ex}$ = 0.2 nm$^{-1}$). This is equivalent to $\alpha'' = -1.35$ nm; therefore, this provides the lower boundary for the elastic cross section: $\alpha_{el} \geq 4\pi|\alpha''|^2 = 23$ nm$^2$. The upper boundary is given by the condition that the modulus of the scattering matrix is equal to or less than unity: $|S_0| \leq 1$. This expression results in the following condition for the real part of the scattering length.

$$\alpha' \leq \frac{1}{2K_{ex}}\left[1 - (1 + 2K_{ex}\alpha'')^2\right].$$

Therefore, $\alpha_{el} = 4\pi\left(|\alpha'|^2 + |\alpha''|^2\right) \leq 30$ nm$^2$. Although these values can vary within the experimental error,[3] we emphasize that the condition $\sigma_{el} \ll \sigma_{inel}$ is satisfied over the temperature range considered in this experiment.

We find the following reports of the collision cross sections. Experimentally, the elastic cross-section is reported to be $\sigma_{el}$ = 0.5 nm$^2$ by measuring the density-dependent blueshift of the paraexciton luminescence. We note, however, that the density was estimated from lineshape analyses, which are inappropriate in the presence of efficient collision-induced loss.[4] For optically accessible 1s orthoexcitons, $\sigma_{el}$ < 30 nm$^2$ is reported from the absence of density-dependent broadening in the luminescence spectra at a given excitation density.[5] Theoretically, quantum Monte Carlo simulations yield $\sigma_{el}$ = 50 nm$^2$ for paraexcitons.[6] We note that the predicted renormalisation of the exciton energy[7] is based on the elastic scattering cross section, ignoring the inelastic scattering process. We believe that such a manifestation is irrelevant for paraexcitons in $Cu_2O$.

**Monte-Carlo simulations of the trapped paraexcitons**

We have conducted Monte-Carlo simulations to confirm that the small condensate fraction results in the enhanced signals in the hot part in spatially resolved luminescence spectra. Here we have adopted the direct simulation Monte-Carlo (DSMC) method[8] that can simulate the gas by tracing the positions and velocities of the excitons, which are rarefied in number. The simulation takes into account all the relevant processes using known parameters: (1) the continuous generation of paraexcitons, (2) lifetime, (3) exciton-phonon interactions, (4) exciton-exciton elastic collisions, (5) exciton-exciton inelastic collisions, (6) trapping excitons by the strain-induced potential, (7) emission of photons and finally (8) inclusion of the condensate. The time step set is 1 ps and we traced the time evolution up to 1000 ns. The lattice temperature is 0.34 K. The details of calculation are the following.

(1) Generation: Paraexcitons having kinetic energy of 12 meV with random directions are generated. The generation rate and the positions the excitons are generated matches our experimental conditions.

(2) Lifetime: paraexcitons are randomly selected and are deleted following the lifetime (300 ns) measured in the present experiment.

(3) Exciton-Phonon interactions: Paraexcitons are scattered by acoustic phonons at rates determined by the deformation potential (2 eV), exciton momenta, and lattice temperature. The magnitudes of the emitted and absorbed phonon momenta are randomly selected so that the selection probability follows the Bose distribution of the phonon bath. The scattering angles of the exciton-phonon collisions are randomly selected.

(4) Exciton-exciton elastic collisions: We calculated the elastic collision rates from the mean free paths that are determined from the s-wave elastic collision cross section and the translational speeds of paraexcitons. When finding collision pairs, we separated the region of interest into many cells and picked exciton pairs randomly out of each cell. The collision angles are randomly selected. We set the scattering length to 0.7 nm. The effect of the rarefaction on the collision rate is properly taken into account.

(5) Exciton-exciton inelastic collisions: We calculated the inelastic collision rates from the spatially dependent densities because the loss rates are independent of the exciton velocity. The selection procedure of the collision pairs is the same as the elastic scattering. In the collision process, one paraexciton of a selected pair is deleted and the other paraexciton is given an additional kinetic energy of 20 meV. This collision process is of crucial importance because even the small number of paraexcitons in the ground state results in an emission of hot paraexcitons due to the small volume of the ground state wave function. The effect of the rarefaction on the loss rate is properly taken into account.

(6) Trapping excitons: We precisely modeled the shape of the three-dimensional trap potential and modified the velocities of the simulated paraexcitons for each time step accordingly.

(7) As we observe photons emitted from the paraexciton gas via the direct emission process, only paraexcitons having the momentum of photons can be detected. We monitored the speed of every simulated paraexctions and recorded their positions when they satisfy the momentum conservation rules.

(8) All excitons are treated as classical particles and no Bosonic scattering process is modeled here. In order to find how the presence of a condensate affects the observed distribution, the condensate fraction is an adjustable parameter here and we kept adding from the thermal part to the ground state (zero kinetic energy) so that the condensate fraction set is maintained all the time.

The simulations reproduce well our experimental results. Supplementary Fig. 4 show the calculated spatially resolved spectra that are expected to emit photons via the direct emission process. As shown the figures, the spectra exhibit the enhancement in the hot exciton part when we have a finite condensate fraction (1%) while keeping the exciton number in the trap almost constant at $2 \times 10^9$ excitons (here the calculation was done by setting each particle representing $5 \times 10^4$ paraexcitons). To understand the enhancement, in Supplementary Video 1 we show the exciton dynamics for the condensate fraction of 1%. Each simulated particle represents $5 \times 10^5$ paraexcitons. Note that hot excitons blowing out of the condensate reach 30−100 μm being cooled by the exciton-phonon interactions. They emit photons at the positions where they have the same momentum as the photon momentum, so that the signal outside the coldest part increases when the occupation of the ground state is finite. This results in the threshold-like increase of the ratio between the high-energy and bottom parts. The condensate (zero translational momentum) cannot be observed directly due to the momentum conservation law in the direct emission process.

The numerical simulation presented here is based on the Graphics Processing Units (GPU)-based parallel computing using a NVIDIA Tesla C2050.

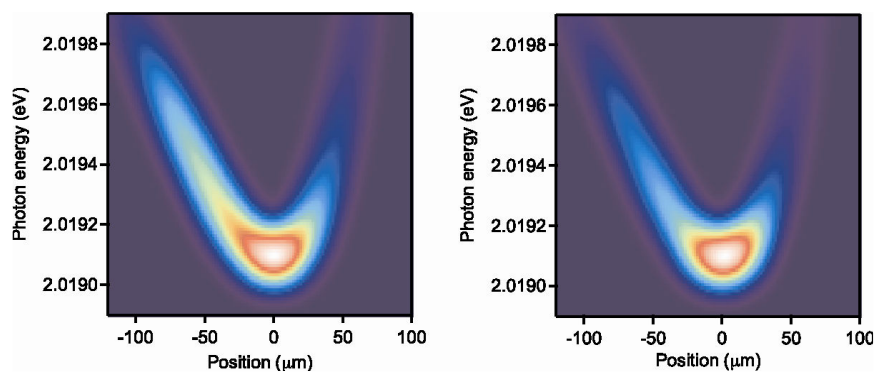

Supplementary Figure 4. Calculated spatially resolved spectra based on the direct simulation Monte-Carlo method. The lattice temperature is set to 0.35 K. Trapped number of paraexcitons is $2 \times 10^9$. (Left) Condensate fraction is 1%. (Right) No condensate. The additional kinetic energy in the inelastic collision loss process is set to 20 meV.

Supplementary Video 1. Calculated dynamics of the rarefied paraexcitons in same the conditions as Supplementary Figure 4 with the condensate fraction of 1%. The spatial distribution is viewed along the X-axis (see Fig. 1a for the dimension). The color roughly designates the translational momenta of paraexcitons. Note the cooling lengths and the positions of the hot paraexcitons that are emitted from the bottom of the trap by the inelastic scattering in the condensate.